\documentclass{an}
\usepackage{graphicx}
\usepackage{times}
\newcommand{\kms}{$\mbox{km s}^{-1}$}
\newcommand{\lsim}{\mathrel{\hbox{\rlap{\hbox{\lower4pt\hbox{$\sim$}}}\hbox{$<$}}}}
\newcommand{\gsim}{\mathrel{\hbox{\rlap{\hbox{\lower4pt\hbox{$\sim$}}}\hbox{$>$}}}}
\newcommand{\sauron}{{\texttt {SAURON}}}
\newcommand{\oasis}{{\texttt {OASIS}}}
\newcommand{\XOasis}{{\small {XOASIS}}}

\newcommand{\farcsec}{\hbox{$.\!\!^{\prime\prime}$}}  

\Volume{00}              
\Year{0000}              
\Month{00}               
\Pagespan{000}{000}      

\begin{document}
\lhead[\thepage]{R. McDermid et al.: \oasis\/ High-Resolution Integral Field Spectroscopy of the \sauron\/
Ellipticals and Lenticulars}
\rhead[Astron. Nachr./AN~{\bf XXX} (200X) X]{\thepage}
\headnote{Astron. Nachr./AN {\bf 32X} (200X) X, XXX--XXX}

\title{\oasis\/ High-Resolution Integral Field Spectroscopy of the \sauron\/
Ellipticals and Lenticulars}

\author{R.\ McDermid$^1$, E.\ Emsellem$^2$, M.\
 Cappellari$^1$, H.\ Kuntschner$^3$, R.\ Bacon$^2$, \\M.\
 Bureau$^4$, Y. Copin$^5$, R.\ L.\ Davies$^{6}$, J.\
 Falc{\' o}n-Barroso$^1$, P.\ Ferruit$^2$, \\D.\ Krajnovi{\' c}$^1$, R.\ F.\
 Peletier$^{7}$, K.\ Shapiro$^1$, F.\ Wernli$^2$, \and P.\ T.\ de Zeeuw$^1$}

\institute{$^1$Sterrewacht Leiden, Niels Bohrweg~2, 
    2333~CA Leiden, The Netherlands\\
$^2$CRAL, 9~Avenue Charles Andr\'e,
    69230 Saint-Genis-Laval, France\\
$^3$Space Telescope European Coordinating Facility, ESO, Karl-Schwarzschild-Str. 2, 85748
  Garching, Germany\\
$^4$Columbia Astrophysics Laboratory, 550~West 120th~Street, 1027 Pupin
    Hall, MC~5247, New York, NY~10027, USA\\
$^5$Institut de Physique Nucl\'eaire de Lyon, 69622 Villeurbanne, France\\
$^6$Denys Wilkinson Building, University of Oxford, Keble Road, Oxford, 
     United Kingdom \\ 
$^7$Kapteyn Institute, Postbus 800, 9700 AV, Groningen, The Netherlands\\
}

\date{Received {October 2003}; 
accepted {November 2003}} 

\abstract{We present a summary of high-spatial resolution follow-up
observations of the elliptical (E) and lenticular (S0) galaxies in the
\sauron\/ survey using the \oasis\/ integral field spectrograph. The
\oasis\/ observations explore the central 8\arcsec$\times$10\arcsec
regions of these galaxies using a spatial sampling four times higher
than \sauron, often revealing previously undiscovered features. Around
75\% (31/48) of the \sauron\/ E/S0s with central velocity dispersion
$\gsim 120$~\kms\/ were observed with \oasis, covering well the
original \sauron\/ representative sample. We present here an overview
of this follow-up survey, and some preliminary results on individual
objects, including a previously unreported counter-rotating core in
NGC\,4382; the decoupled stellar and gas velocity fields of NGC\,2768;
and the strong age gradient towards the centre of NGC\,3489.
 }
\correspondence{mcdermid@strw.leidenuniv.nl}

\maketitle

\section{Introduction}

The \sauron\/ project (de Zeeuw et al. 2002) is a study of the
structure of 72 representative nearby early-type galaxies and spiral
bulges based on measurements of the two-dimensional stellar kinematics
and distribution of absorption line-strengths, together with the
kinematics and distribution of ionized-gas, measured with \sauron, a
panoramic integral-field spectrograph (IFS) mounted at the William
Herschel Telescope (WHT), La Palma (Bacon et al. 2001). The aims of
this survey and the observed galaxy sample are described in de Zeeuw et
al. (2002).  The stellar kinematic maps for the sub-sample of 48 E/S0s
is presented in Emsellem et al. (2003).

To provide a large field of view, the spatial sampling of \sauron\/ was
set to 0\farcsec94$\times$0\farcsec94 (per lenslet) for the survey,
therefore often undersampling the typical seeing at La Palma
(0\farcsec8 FWHM). This does not, in principle, affect the results for
the main-body of early-type galaxies, as they generally exhibit
smoothly-varying structure on scales larger than the seeing. Towards
the galaxy nucleus, however, there are often sharp, localised features
in the kinematics, such as decoupled cores or central disks, as well as
distinct stellar populations and ionized-gas distributions. Such
features may be only partially resolved in the \sauron\/ data, and in
some cases are not even visible.

In addition, at HST resolution, all elliptical galaxies exhibit
power-law central luminosity profiles: $\Sigma (r) \propto
r^{-\gamma}$, with clear trends between the nuclear luminosity profiles
and certain global properties, such as the degree of rotational
support, isophotal shape, and stellar populations. It is therefore
crucial to fill the gap between the medium (few 100s of pc) to
large-scale (few kpc) structures probed with \sauron\/ and the inner
($< 200$~pc) components probed by HST. We have thus conducted a
complimentary study on a subset of the \sauron\/ sample using the
\oasis\/ (Optically Adaptive System for Imaging Spectroscopy)
spectrograph, mounted on the Canada-France-Hawaii Telescope (CFHT),
Hawaii.

\section{The \oasis\/ Spectrograph}

\oasis\/ is an IFS based on the TIGER concept (Bacon et al. 1995) and
is designed for high spatial resolution observations. It is a
multi-mode instrument, with both imaging and IFS capabilities, and can
be assisted by an adaptive optics (AO) system.  \oasis\/ operated at
the Cassegrain focus of the CFHT since 1997 (with the PUEO AO system),
but was recently transferred to the Nasmyth focus of the WHT behind the
NAOMI AO system (March 2003).  All observations presented here were
obtained at the CFHT, hence in the following we solely refer to the
\oasis\/ CFHT configurations.

The imaging mode of \oasis\/ is used primarily for accurate target
acquisition. There is a selection of gratings and filters within
\oasis, giving low and medium spectral resolution modes within the
0.43~$\mu$m to 1~$\mu$m wavelength range. Via the use of different
enlargers, there is also a range of spatial sampling, from
0\farcsec4$\times$0\farcsec4 lenslets covering a
15\arcsec$\times$12\arcsec\ field, down to
0\farcsec04$\times$0\farcsec04 sampling across a
1\farcsec5$\times$1\farcsec2 field, for use with adaptive optics. Since
most of the objects in the \sauron\/ sample of E/S0s do not have a
bright enough nearby guiding source ($m_V < 16$), we decided to use the
$f/8$ (no AO) mode of \oasis\/. We still chose a rather small spatial
scale of 0\farcsec27$\times$0\farcsec27 per lenslet for the
observations to properly sample the generally excellent seeing at Mauna
Kea, providing over 1000 individual spectra in a
10\arcsec$\times$8\arcsec\ field-of-view.

As a complimentary data-set to the \sauron\/ survey, \oasis\/ was
configured to give similar spectral coverage and resolution as \sauron,
resulting in a wavelength range of $4760$--$5558$~\AA, with a resolution of
4.2~\AA\/ FWHM sampled at $1.95$~\AA~pix$^{-1}$. This configuration
is suitable for measuring stellar kinematics in early-type galaxies,
and also covers key absorption/emission features such as Mg\,$b$,
H$\beta$, [O{\small III}]$\lambda\lambda$4959,5007, and a number of Fe
lines.

\section{Observed Sample and Data Reduction}

\subsection{Observed Sample}

\begin{figure}
\resizebox{\hsize}{!}
{\includegraphics[angle=90]{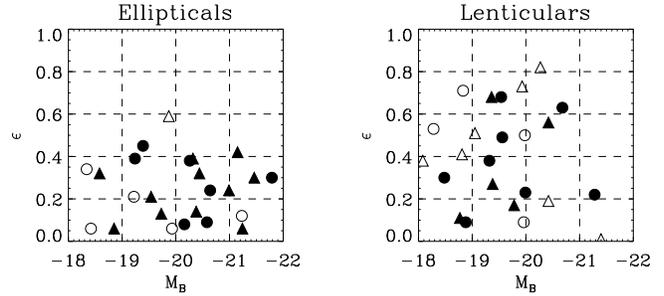}}
\caption{{\small \sauron\/ survey sample of ellipticals and
  lenticulars. Objects are separated as `field' (circular symbols) and
  `cluster' (triangular symbols). Filled symbols indicate the subset of
  objects also observed with \oasis.}\vspace{-12pt}}
\label{fig:sample}
\end{figure}

From the representative sample of the \sauron\/ survey, objects with
central velocity dispersion less than the instrumental dispersion of
\oasis\/ ($\sim 120~$\kms) were excluded, removing almost all the
early-type spiral bulges from the sample, and six of the 48 \sauron\/
E/S0 galaxies. To ensure a homogeneous sample, observational efforts
were thus concentrated on the E/S0s of the \sauron\/ survey. Of the
remaining 42 galaxies, a total of 31 galaxies were observed with
\oasis\/ during three observing runs between March 2001 and April
2002. Figure \ref{fig:sample} presents the final observed \oasis\/
sub-sample of the E/S0s of the \sauron\/ survey, and shows that the
plane of ellipticity versus absolute $B$-magnitude is covered quite
homogeneously by the \oasis\/ sub-sample.

\subsection{Data Reduction and Binning}

The data were reduced using the publicly available \XOasis\/
software (Rousset 1992) developed at CRAL (Lyon). The reduction steps
include bias and dark subtraction, extraction of the spectra using an
instrument model, wavelength calibration, low-frequency
flat-fielding, removal of cosmic rays, sky subtraction, and
flux-calibration. Galaxy observations were composed of two or more
exposures, each offset by some small, non-integer number of lenslets to
provide oversampling and avoid systematic effects
due to bad CCD regions. Multiple exposures were combined by first
truncating to a common wavelength region and centering the spatial
coordinates on the galaxy nucleus via reconstructed images. Exposures
were then re-normalised to account for transparency variations, and
resampled onto a common spatial grid of
0\farcsec2$\times$0\farcsec2. Co-spatial spectra were then combined via
an optimal summing routine, taking into account the error spectra which
are propagated through the reduction.

In order to provide reliable, unbiased measurements of stellar
kinematic parameters, as well as absorption line strengths, we bin all
data to a minimum signal-to-noise ratio ($S/N$) of 60~pixel$^{-1}$
using the Voronoi 2D-binning developed by Cappellari \&
Copin (2003).


\section{Data Analysis}

\subsection{Stellar Kinematics}

Similarly to Emsellem et al. (2003), stellar absorption-line kinematics
were derived for the galaxies by directly fitting the spectra in
pixel-space. This method was chosen over Fourier-based methods due to
its robustness to contamination by nebular emission lines, which can
often be strong in the central regions of early-type galaxies. Using
this `pixel-fitting' method (PXF; e.g., van der Marel 1994), emission
lines are simply excluded from the fitting process, and only
information from emission-free spectral regions is used.

\begin{figure}
\resizebox{\hsize}{!}
{\includegraphics[angle=0]{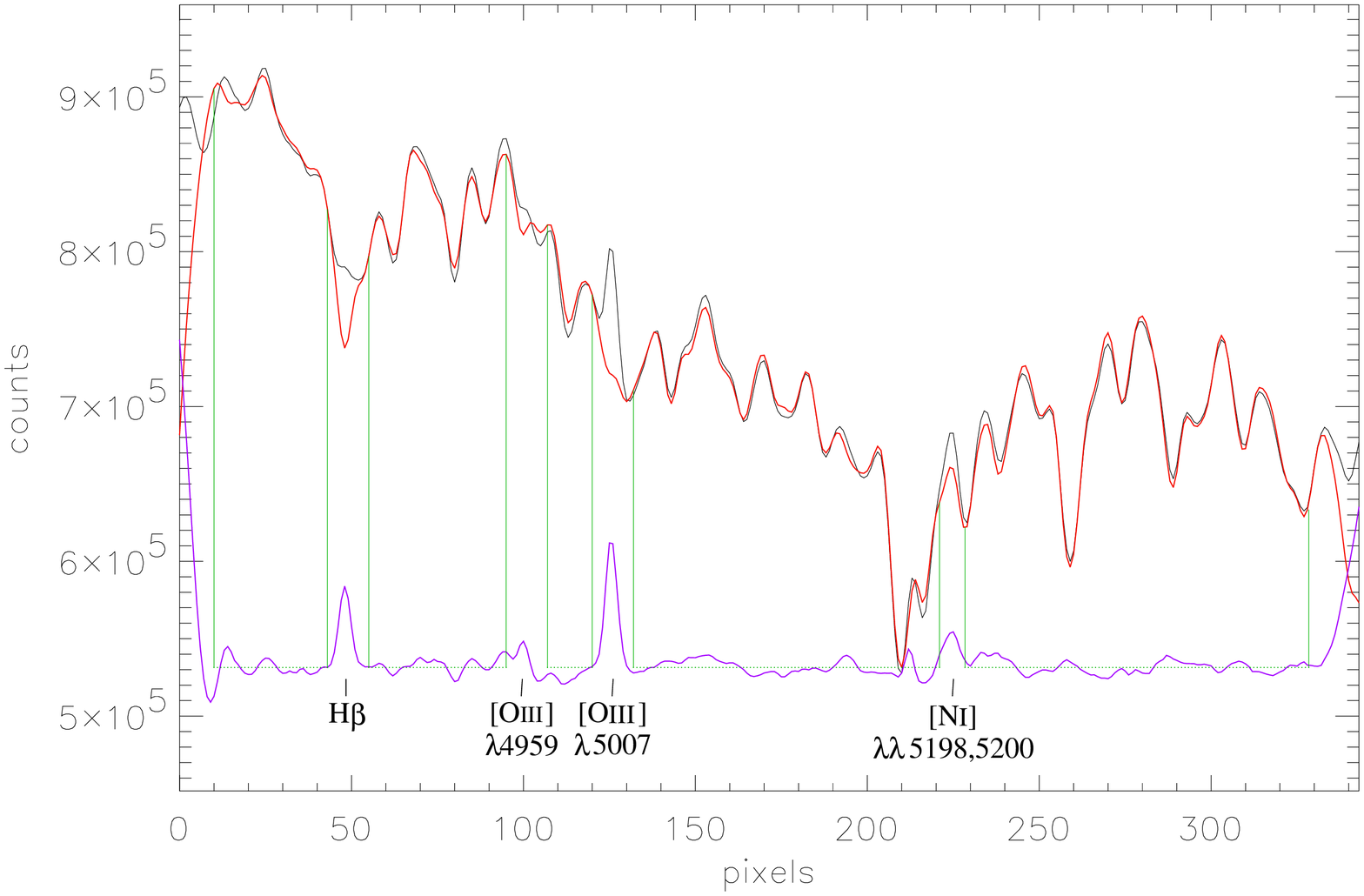}}
\caption{{\small Example of optimal template fit using the PXF method, for a
  spectrum of NGC\,2768. The lower spectrum shows the residual emission
  lines after the template fit, which are then fitted using single
  Gaussians to obtain the gas properties. Vertical lines show regions
  around the emission which are excluded from the fit.}\vspace{-10pt}}
\label{fig:opt_temp}
\end{figure}

This method can, however, be sensitive to template mismatch effects,
which can bias the result. This problem was minimized by selecting an
`optimal template' at each iteration of the fitting process. For each
trial set of kinematic parameters, an optimal linear combination of
appropriately convolved absorption spectra were fitted to the data
(Figure \ref{fig:opt_temp}). The
library of absorption spectra used were taken from the single-burst
stellar population (SSP) models of Vazdekis (1999), with the addition
of several individual stellar spectra with strong Mg\,$b$ to compensate
for the near-solar abundance ratio inherent in the SSP models.

\subsection{Gas Properties}

As a natural by-product of the PXF method employed for the stellar
kinematics, the derived optimal template can be used to separate the
absorption and emission line components of the data. By subtracting the
optimal template, one obtains a residual spectrum in which the significant
emission line feature are revealed. We then determine the distribution of
the emission features, and measure the kinematics of the ionized-gas,
by fitting the emission line profiles of these continuum-free spectra
with a simple Gaussian.

\subsection{Line Strengths}

The \oasis\/ spectral range contains a number of key absorption
features which can be used as diagnostic tools to determine the
distribution of stellar populations within a galaxy, based on
measurements of their age and metallicity. Many of the absorption
features in this range can also be significantly altered due to
emission features. Therefore, the emission spectrum models (in this
case, simple Gaussians) fitted to the residual of the optimal template
are subtracted from the original data before measuring the absorption
line strengths. Finally, the absorption line strengths are calibrated
onto the well-established LICK/IDS system (e.g. Trager et al. 1998).

\section{Preliminary Results}


\subsection{Decoupled Core in NGC\,4382}

\begin{figure*}
\resizebox{15.0cm}{!}
{\includegraphics[]{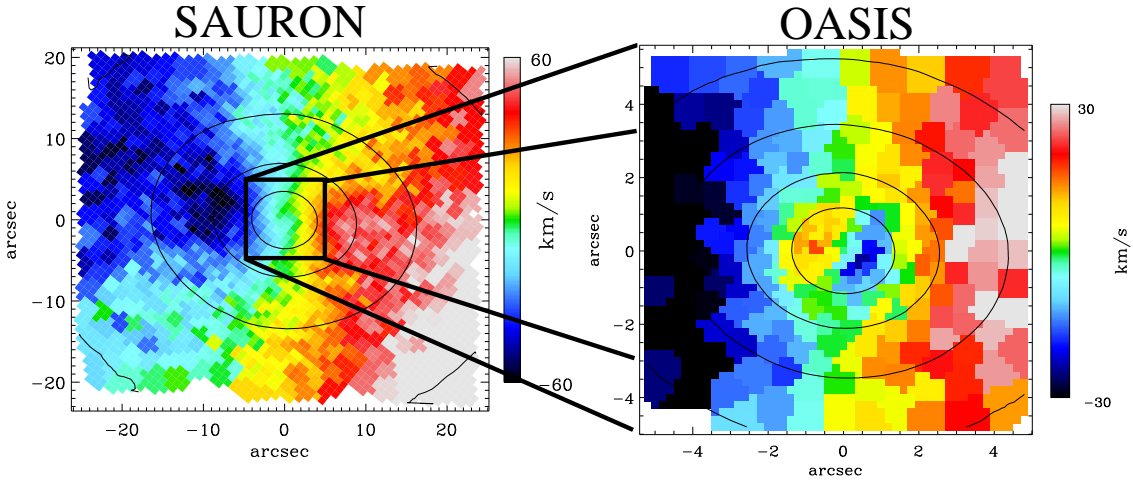}}
\caption{{\small Voronoi 2D-binned \sauron\/ velocity field of NGC\,4382 (left
  map) with a box showing the outline of the \oasis\/ field (right
  map). The zero-velocity level is indicated by the colour green on
  both maps. Isophotes are overplotted to show the luminosity
  distribution. The \oasis\/ data clearly reveal the central
  counter-rotating KDC.}}
\label{fig:ngc4382_kdc}
\end{figure*}

Figure \ref{fig:ngc4382_kdc} shows an example of how the \oasis\/ data
can be used to reveal central features of galaxies in the \sauron\/
survey. The left panel of this figure presents the \sauron\/ velocity
field of NGC\,4382. There is a low-level `kink' in the zero-velocity
contour as it passes through the galaxy centre. The \oasis\/ data
(right panel) clearly reveal this as a counter-rotating kinematically
decoupled component (KDC), which appears almost aligned with the
galaxy's main body. This KDC was not previously reported in the
literature.

\subsection{Decoupled Gas and Stellar Kinematics in NGC\,2768}

\begin{figure*}
\resizebox{15.0cm}{!}
{\includegraphics[]{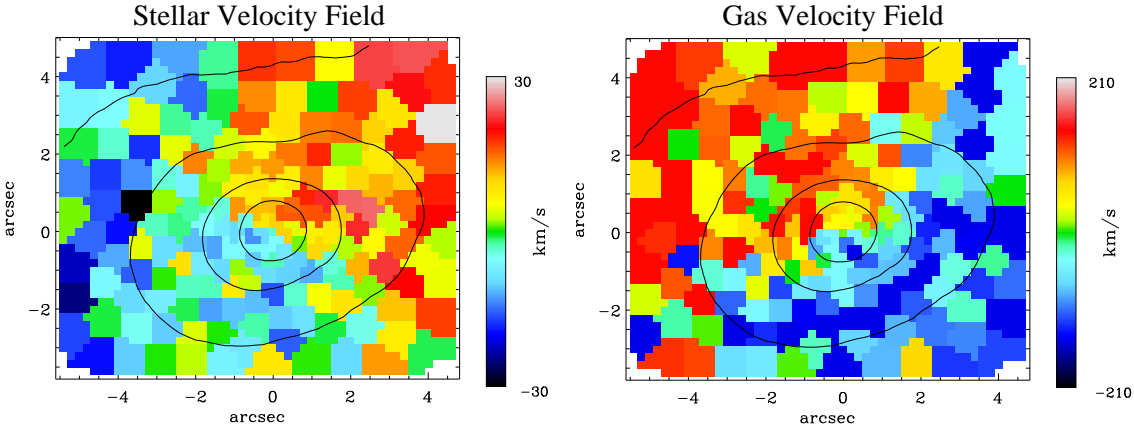}}
\caption{{\small \oasis\/ stellar (left) and ionized-gas (right) velocity
  fields for the polar-ring galaxy NGC\,2768. The decoupled rotation
  of the stars and gas can clearly be seen, with the two components
  rotating around near-orthogonal axes. Isophotes from the
  reconstructed image are overplotted, showing the total flux within
  each \oasis\/ spectrum. Distortion of these isophotes indicates dust
  features.}}
\label{fig:ngc2768_star_gas}
\end{figure*}

Figure \ref{fig:ngc2768_star_gas} presents the stellar (left panel) and
gas (right panel) velocity fields for the polar-ring galaxy
NGC\,2768. The stellar component rotates around the apparent short-axis
of the galaxy. The gas, however, rotates around the apparent long-axis,
perpendicular to the stars. This illustrates how we can separate the
stellar and gas properties, using the optimal template fit. There is
some additional structure in the upper regions of the stellar velocity
field, likely due to dust extinction.

\subsection{Post-Starburst Galaxy, NGC\,3489}

\begin{figure*}
\resizebox{15.0cm}{!}
{\includegraphics[]{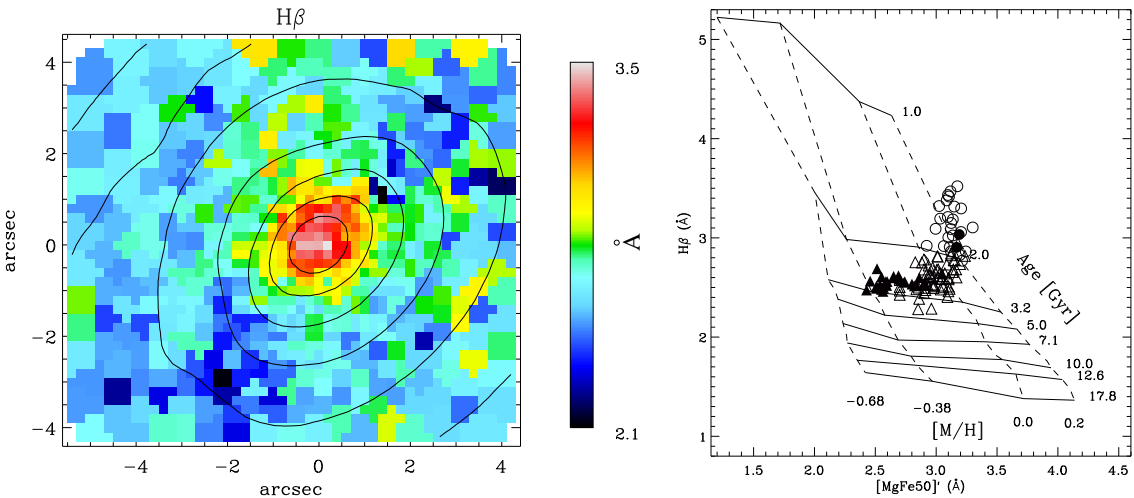}}
\caption{{\small Left panel: Map of H$\beta$ absorption line strength in NGC\,3489,
  measured {\it after} subtraction of any H$\beta$ emission features
  which may be present. The steep rise in H$\beta$ absorption is still
  present without subtracting the emission component, implying that
  this central peak is not an artefact of the emission correction. Right
  panel: H$\beta$ absorption strength vs the abundance-insensitive
  [MgFe50]$^\prime$ metallicity index overplotted with a grid of
  stellar population models from Vazdekis (1999). Open symbols
  represent \oasis\/ measurements; filled symbols represent \sauron\/
  measurements of the same galaxy, binned in 1\arcsec\ circular
  annuli. Circles indicate measurements inside a 1\arcsec\ radius of
  the centre; triangular symbols indicate measurements outside this
  radius: up to $\sim 5$\arcsec\ radius for \oasis, and $\sim
  20$\arcsec\ radius for \sauron.}}
\label{fig:ngc3489_stellarpop}
\end{figure*}

Figure \ref{fig:ngc3489_stellarpop} presents a map of H$\beta$
absorption strength for the galaxy NGC\,3489 (left panel) which shows a
pronounced peak in the central 1\arcsec, indicative of a young central
stellar population. The right panel of Figure
\ref{fig:ngc3489_stellarpop} quantifies this, plotting H$\beta$
absorption strength against the abundance-insensitive metallicity
indicator [MgFe50]$^\prime$\footnote{[MgFe50]$^\prime = \frac{0.45
\times \mathrm{Mg}b + \mathrm{Fe5015}}{2}$ (Kuntschner et al. in
prep.)} from the \oasis\/ data. The young stellar population in the
core of this galaxy indicates that it is in a post-starburst phase,
with a luminosity-weighted age of around 1.5~Gyr. Equivalent \sauron\/
data are also shown, illustrating that both data sets are consistent
(see caption for details).

\section{Summary and Future Plans}

We have conducted extensive high-spatial resolution
observations of the E/S0 sub-sample of the \sauron\/ survey using the
\oasis\/ IFS. This provides a unique data set on the centres of these
galaxies, complimenting the panoramic view delivered by \sauron. We
have presented here some preliminary results to illustrate the quality
of the data and the analysis techniques we use. The \oasis\/ data set
will be published as a complimentary project to \sauron, relating the
inner properties of these galaxies with those of the outer regions, as
well as providing a high-resolution catalogue of a substantial sample
of early-type galaxies. The data will be made publicly available when
the final reduction and analysis is complete.

\acknowledgements It is a pleasure to thank 
Pierre Martin for enthusiastic
 support
throughout this observational campaign.

\end{document}